\documentclass[10pt,twocolumn,pra,aps,showpacs,superscriptaddress,floatfix]{revtex4-1}
\usepackage[utf8]{inputenc}
\usepackage{bm}
\usepackage[usenames]{color}
\usepackage{multirow}
\usepackage{amssymb}
\usepackage{amsbsy}
\usepackage{amsmath}
\usepackage{stmaryrd}
\usepackage{graphicx}
\usepackage{epsfig}
\usepackage{placeins}
\usepackage{ulem}
\usepackage{hyperref}
\usepackage{tikz}
\usepackage{pgf}
\usepackage{braket}
\hypersetup{colorlinks=true,
citecolor=blue,
linkcolor=red,
urlcolor=black,
pdfmenubar=true}
\usepackage[textsize=tiny]{todonotes}

\makeatletter
\newcommand{\makecomment}[2]{\@ifundefined{@captype}{\todo[noinline,color=#1]{#2}}{\todo[inline,color=#1]{#2}}}
\def\InFloat{\ifnum\@floatpenalty<0 inline\else noinline\fi}
\makeatother

\usepackage{relsize}

\makeatletter

\newcommand{\be}{\begin{equation}}
\newcommand{\ba}{\begin{align}}
\newcommand{\ee}{\end{equation}}
\newcommand{\ea}{\end{align}}

\renewcommand{\perp}{\scriptscriptstyle{\perp}}

\newcommand{\hc}{\mathrm{h.c.}}

\newcommand{\etal}{{\it et al.}}

\begin{document}
\title{Topological charge pumping in the interacting bosonic Rice-Mele model}

\author{A. Hayward}
\affiliation{Institute for Theoretical Physics, Georg-August-Universit\"at G\"ottingen, Friedrich-Hund-Platz 1,
37077 G\"ottingen, Germany}
\affiliation{Arnold Sommerfeld Center for Theoretical Physics, Ludwig-Maximilians-Universit\"at M\"unchen, 80333 M\"unchen, Germany}
\affiliation{Fakult\"at f\"ur Physik, Ludwig-Maximilians-Universit\"at, Schellingstrasse 4, 80799 M\"unchen, Germany}

\author{C. Schweizer}
\affiliation{Fakult\"at f\"ur Physik, Ludwig-Maximilians-Universit\"at, Schellingstrasse 4, 80799 M\"unchen, Germany}
\affiliation{Max-Planck-Institut f\"ur Quantenoptik, Hans-Kopfermann-Strasse 1, 85748 Garching, Germany}

\author{M. Lohse}
\affiliation{Fakult\"at f\"ur Physik, Ludwig-Maximilians-Universit\"at, Schellingstrasse 4, 80799 M\"unchen, Germany}
\affiliation{Max-Planck-Institut f\"ur Quantenoptik, Hans-Kopfermann-Strasse 1, 85748 Garching, Germany}

\author{M. Aidelsburger}
\affiliation{Fakult\"at f\"ur Physik, Ludwig-Maximilians-Universit\"at, Schellingstrasse 4, 80799 M\"unchen, Germany}
\affiliation{Max-Planck-Institut f\"ur Quantenoptik, Hans-Kopfermann-Strasse 1, 85748 Garching, Germany}

\author{F. Heidrich-Meisner}
\affiliation{Institute for Theoretical Physics, Georg-August-Universit\"at G\"ottingen, Friedrich-Hund-Platz 1,
37077 G\"ottingen, Germany}

\date{\today}

\begin{abstract}
  We investigate topological charge pumping in a system of
  interacting bosons in the tight-binding limit, described by
  the Rice-Mele model. 
 An appropriate topological invariant for the many-body
  case is the change of polarization per pump cycle, which we compute for
  various interaction strengths from infinite-size matrix-product-state
  simulations. 
We verify that the charge pumping remains quantized as
  long as the pump cycle avoids the superfluid phase.
   In the limit of hardcore bosons, the
  quantized pumped charge can be understood from single-particle properties
  such as the integrated Berry curvature constructed from Bloch states, while
  this picture breaks down at finite interaction strengths. These
  two properties -- robust quantized charge transport in an interacting system of bosons
  and the breakdown of a single-particle invariant -- could both be measured with
  ultracold quantum gases extending a previous experiment [Lohse~\etal,  Nature
  Phys. {\bf 12}, 350 (2016)].  Furthermore, we
  investigate the entanglement spectrum of the Rice-Mele model and argue that
  the quantized charge pumping is encoded in a winding of the spectral flow in
the entanglement spectrum over a pump cycle. 
\end{abstract}

\maketitle

\section{Introduction}
\label{sec:intro}

The experimental and theoretical investigation of topological states of matter
is a key topic in condensed matter physics \cite{rachel2018interacting, Qi2011, Hasan2010}, ultracold quantum gases \cite{Cooper2018, aidelsburger2018artificial, Galitski2013} and photonics \cite{LeHur2016, ozawa2018topological}. 
While the theoretical classification of noninteracting symmetry-protected topological states is
complete \cite{Altland1997,Schnyder2008}, the investigation of topology
beyond noninteracting particles, zero temperature and closed quantum systems is an active field of research. 
In recent years, different characterizations have been developed, showing that only certain aspects of topology can survive in mixed states (see, e.g., \cite{rivas2013density,Bardyn2013,Linzner2016,Zheng2018,Grusdt2017,Bardyn2018}) or nonequilibrium situations \cite{Caio2015,Wang2016,Hu2016,Budich2016,Wilson2016,Wang2017,Qin2018}.
However, for interacting systems at zero temperature, generalizations of topological invariants have been identified \cite{Wang2010,Gurarie2011, Wang2012,Wang2013,rivas2013density, shen2014hall, huang2014topological,budich2015topology, Zheng2018},
and it was shown that for many examples, such as Chern insulators, the topology is  preserved for sufficiently small interactions~ (see, e.g., \cite{Thouless1982,niu1984quantised,Varney2011,Wang2012,turner2013beyond, Yoshida2014}).

Experiments with ultracold atomic gases are now able to explore aspects of topological physics
providing direct access to non-trivial quantities such as Chern numbers \cite{Aidelsburger2015,Wu2016,Asteria2018,tarnowski2017characterizing}, 
Berry phases, Berry curvatures \cite{Jotzu2014, Duca2015, Flaeschner2016, tarnowski2017characterizing}, 
the dynamics of edge states~\cite{Atala2014, Mancini2015, Stuhl2015} 
or the topological properties of one-dimensional (1D) symmetry-protected topological phases such as the Su-Schrieffer-Heeger model (SSH) \cite{Atala2013,Meier2016,Leder2016}. 
Moreover, some of the most paradigmatic two-dimensional (2D) lattice models with topological band-structures were implemented, including the 
Hofstadter  model~\cite{Aidelsburger2013,Miyake2013,kennedy2015observation} and variants \cite{Atala2014,Stuhl2015,Mancini2015,Tai2017,An2017} and the Haldane model~\cite{Jotzu2014,Asteria2018}.
Most of these experiments, however, focused on noninteracting systems and there is a strong ambition to go beyond single-particle dynamics and to investigate topological phenomena in genuine
many-body systems. 

Here, we consider the effects of interactions in a topological charge-pumping set-up that is realizable in quantum-gas experiments. Topological charge pumps were introduced in the seminal work by Thouless ~\cite{thouless1983quantization}.  
In the noninteracting case it is well known that an adiabatic cyclic evolution in the space of Hamiltonian parameters 
leads to a quantized charge transport per cycle.
In fact, this quantization is intimately related to the quantized Hall conductivity of a (2D) Chern insulator, in the sense that the cyclic parameter of the 1D Hamiltonian (that describes the charge pump) corresponds to   a 
quasi-momentum in the 2D model \cite{thouless1983quantization, niu1984quantised}.  
Experiments on topological Thouless pumping were performed with ultracold atoms \cite{Lohse2016,Nakajima2016,Schweizer2016,Lu2016} and photons \cite{Kraus2012, verbin2015topological} and have recently been extended to higher dimensions \cite{Lohse2018, Zilberberg2018}.

Interacting charge pumps have been studied theoretically, 
both for  fermions \cite{requist2017approximate, nakagawa2018breakdown} and bosons \cite{berg2011quantized}.
Originally, it was shown by Thouless that quantization is unaffected by weak interactions under fairly broad assumptions \cite{niu1984quantised}. 
However, numerical simulations have shown that strong interactions can lead to a breakdown of the quantized pumping by closing the many-body gap \cite{nakagawa2018breakdown}.
Moreover, charge pumps are convenient tools to characterize the topology of interacting many-body
systems in numerical simulations (see, e.g., \cite{zaletel2014flux, grushin2015characterization,Grusdt2014}).

In the limit of hardcore bosons, the interacting 1D charge pump, as reported in the experimental work
Ref.~\cite{Lohse2016}, has a simple interpretation. The  model can be mapped
onto noninteracting spinless fermions. For a completely filled band of
fermions, all quasi-momentum components are homogeneously populated and the
total amount of charge transported per cycle is determined by the sum of the
Berry curvature over all quasi-momenta. In this work we are interested in the
regime of finite interaction strengths, where this mapping is no longer valid. Finite
interaction strengths result in an inhomogeneous momentum distribution and a many-body
characterization of the transport via the polarization needs to be used. 

Let us give a more formal account of the relevant topological invariants. To quantify a charge pump, one can consider the
evolution of the many-body polarization $P(t)$ of a state $\ket{\Psi(t)}$
over the course of the adiabatic driving of a time-dependent Hamiltonian
$H(t)$. 

The modern theory of polarization associates $P$ with the mean position of the
charge distribution per unit cell in the case of translationally invariant states~\cite{Resta1994}.  
For a system of length $La$ with periodic boundary conditions, 
one can define the polarization for a many-body wave function $\ket{\Psi(t)}$ via \cite{Resta1998}:
\begin{equation}
  \label{eq:pol-MB1}
  P(t)=\frac{qa}{2\pi}\operatorname{Im}\ln\left\langle\Psi(t) \middle|
  e^{\frac{i2\pi}{La}\hat{X}} \middle| \Psi(t)\right\rangle\quad(\bmod \,qa), 
\end{equation}
where $\hat{X} =\sum_{x=0}^{L-1} a \left(x-x_{0}\right)\hat{n}_{x}$ is the
position operator, $\hat n_x$ is the local density operator,   $q$ is the charge per particle, $a$
is the length of the unit cell, $L$ is the number of sites, and $x_0$ is the unit-cell center. 
Importantly, $P$ is only defined modulo $qa$ and it has the units of a dipole moment. 
We keep $q$ in the equations for clarity, yet for our case of neutral atoms, $q=1$.

In general, the total transported charge $\Delta Q$ can  be related to the polarization via:
\begin{equation}
  \label{eq:int-current}
  \Delta Q  =  \int _ { 0 } ^ { \Delta t } d t J ( t ) = \frac{1}{a}\int_0^{\Delta t}dt\,  \partial_tP(t),
\end{equation}
where $J(t)$ is the current density of the system.

Now we consider the case of charge pumps which are described by time-periodic Hamiltonians, such that $H(t+T) = H(t)$, with $T$ the period of the pump cycle. 
Assuming perfect adiabaticity,  the polarization is  cyclic in $T$ as well, with $P(T) \mod qa = P(0)$. The final
expression in Eq.~\eqref{eq:int-current} with $\Delta t = T$ is the winding number of
the polarization, which implies the quantization of the pumped charge for each cycle.

\begin{figure}[t]
   \centering
        \includegraphics{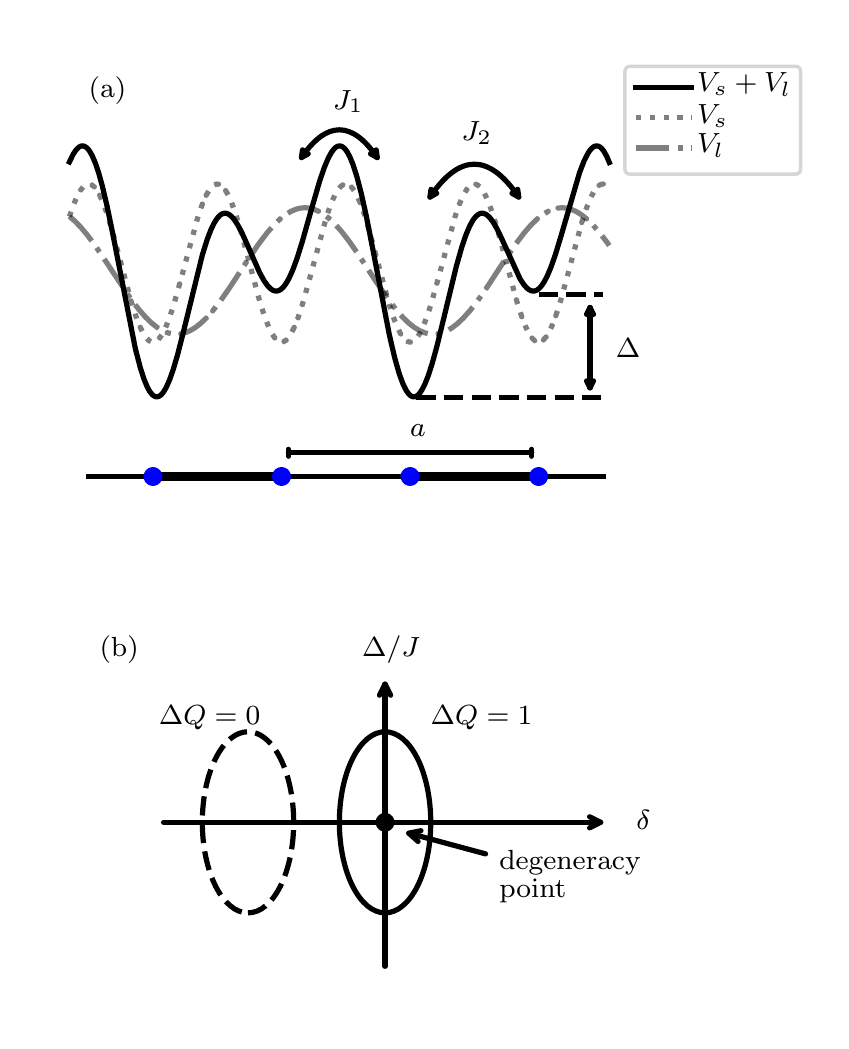}
        \caption{(Color online) {\it Schematic of the Rice-Mele model and of exemplary pump cylces}: (a)  A bichromatic optical potential created from two optical standing waves with wavelength $\lambda_l = 2 \lambda_s$ can be described by a tight-binding lattice model, the Rice-Mele model, with staggered potential $\Delta$ and alternating tunneling strength  $J_{i}=J\,\bigl(1 + (-1)^i \delta\bigr)$, where $i=\{1,2\}$. $\delta$ is the dimensionless dimerization parameter [Eq.~(\ref{eq:model})] and $a = \lambda_l/2$ is the length of the unit cell. (b) These parameters can be tuned by changing the lattice potentials $V_{l,s}$, and the  relative phase $\phi$ of the two optical potentials [Eq.~(\ref{eq:superlattice})]. Periodic tuning of these parameters defines a pump path.
        The pumped charge $\Delta Q$ of a pump path can be either $\Delta Q=1$ (solid line) or $\Delta Q=0$ (dashed line) depending on whether the pump cycle encloses the degeneracy point at $\delta=\Delta=0$ or not. }  \label{fig:model}
\end{figure}

The expression for the many-body polarization reduces to the usual form 
for noninteracting fermions for a filled band \cite{resta1992theory,king1993theory, Resta1994}:
\begin{equation}
\label{eq:pol-fermi}
P_{\mathrm{NI}}(t) = \frac{-iq}{2\pi}\int_{-\pi/a}^{\pi/a} dk \braket{u(k,t)|\partial_k|u(k,t)},
\end{equation}
where $\ket{u(k,t)}$ are single-particle momentum-eigenstates of the system and $k \in [-\pi/a, \pi/a)$ is the quasi-momentum.
In this case, the polarization is  proportional to the Zak phase associated with a filled band,
which is defined modulo $2\pi$ \cite{zak1989berry}. The charge pumped per cycle can then be expressed as:
\begin{equation}
  \label{eq:chern-number}
\Delta Q = \frac{q}{2\pi}\int^{\pi/a}_{-\pi/a} dk \int_0^T dt \,\Omega(t,k) = q\nu\,, \end{equation}
where  $\nu$ is the Chern number associated with the Berry curvature of the pump cycle:
\begin{equation} 
  \label{eq:berry-curvature}
\Omega(k,t) =
i\braket{\partial_k u(k,t)|\partial_t u(k,t)} - c.c. \,,
\end{equation}
defined on a torus of dimension $T\times2\pi/a$.
If Eq.~\eqref{eq:chern-number}, with a suitable choice of states $|u(k,t)\rangle$, yields the actual quantized charge in a many-body system, then we say that
a single-particle description applies.

In this paper, we investigate interacting bosonic charge pumps in the tight-binding Rice-Mele model \cite{rice1982elementary}, which consists of alternating hopping-matrix elements and alternating on-site potentials [see Fig.~\ref{fig:model}(a)]. In order to model a charge pump, both parameters are varied periodically [see Fig.~\ref{fig:model}(b)]. We pursue two main goals: First, we reveal the breakdown of the single-particle interpretation in a parameter regime, where the charge pump remains topologically protected away from the limit of hardcore bosons. Measuring both the quasi-momentum distribution function and the center-of-mass motion of the cloud as a measure of 
the quantization of charge pumping  are sufficient to demonstrate this breakdown in an experiment. Both observables are readily available in existing cold-atom experiments. Realizing this regime would constitute a nontrivial way of observing topologically protected charge pumping in an interacting system and may be an important step towards experimental studies of topological effects in many-body systems. Second, we investigate entanglement properties and observe that the quantized charge pump is reflected in a winding of the spectral flow in the entanglement spectrum.

We use matrix-product-state (MPS) methods in an infinite-system size formulation (iDMRG)~\cite{mcculloch2008infinite, Vidal2007, Schollwoeck2011} to establish that the interacting topological charge pump remains quantized as long as the many-body gap does not close along the cycle for sufficiently strongly-interacting bosons. As the system enters the superfluid region, correlations in the system diverge, and the transported charge is no longer quantized. Therefore, we focus on the Mott-insulating phase, where we explicitly demonstrate that there exists a parameter regime, in which an interpretation in terms of single-particle states results in a non-quantized transport, while the true transported charge actually remains protected and quantized.  

We further relate the quantized charge pumping to the entanglement spectrum of the
reduced density matrix in spatial bipartitions.  The relationship between the
entanglement spectrum, edge modes, and topological properties is well
documented (see, e.g., \cite{li2008entanglement, fidkowski2010entanglement,
pollmann2010entanglement,qi2012general}).  In particular, Li and Haldane were able to
show the correspondence between the low-energy structure of the entanglement
spectrum of fractional quantum Hall states and their topology
\cite{li2008entanglement}.  In 1D systems, symmetry-protected
topological states \cite{chen2011complete, schuch2011classifying} imply an exact degeneracy in the entanglement spectrum
\cite{pollmann2012symmetry}.  Indeed,
we show that charge pumping is encoded in a parity protection which appears for
strong interactions.  This results in a winding of the spectral flow over the pump
cycle.

The paper is organized as follows. We introduce the interacting bosonic Rice-Mele model and its phase diagram in Sec.~\ref{sec:model}, where we also present the iDMRG method. 
In Sec.~\ref{sec:pump} we introduce the pump cycles studied in this work and show how charge pumping remains quantized in the presence of interactions via the winding of polarization.
In Sec.~\ref{sec:mdf} we discuss the one-body correlation functions 
and their associated quasi-momentum distributions as possible experimental observables, 
and demonstrate the breakdown of a single-particle interpretation of quantized
charge pumping away from the limit of hardcore bosons.  
Aspects relevant for experimental realizations are discussed in Sec.~\ref{sec:exp}.  
In Sec.~\ref{sec:ent}, we relate the quantized pumping to the spectral flow in the entanglement spectrum.
We conclude with a summary and an outlook in Sec.~\ref{sec:sum}.


\section{Interacting Rice-Mele model}\label{sec:model}
\subsection{Realization and Hamiltonian}
The Rice-Mele model~\cite{rice1982elementary} is a 1D
tight-binding lattice model with alternating nearest-neighbor hopping-matrix elements $J_{i}=J\,\bigl(1 + (-1)^i \delta\bigr)$, $\delta \in [-1,1]$, and a staggered on-site
potential with strength $\Delta$. In typical cold-atom experiments \cite{Lohse2016,Nakajima2016}, it can be realized with a bichromatic optical lattice, which consists of two optical lattices with wavelength $\lambda_s$ and $\lambda_l=2\lambda_s$ as illustrated in Fig.~\ref{fig:model}(a). The corresponding optical potential is of the form 
\begin{equation}
    V(x) = V_l \sin^2\left(\frac{\pi x}{\lambda_l}- \frac{\phi}{2}\right) + V_s \sin^2\left(\frac{\pi x}{\lambda_s} + \frac{\pi}{2} \right)\,,
    \label{eq:superlattice}
\end{equation}
where $V_{s,l}$ are the respective potential depths and $\phi$ is the relative phase between the two potentials. For sufficiently deep lattices the physics can be described by the Rice-Mele tight-binding lattice model, which for bosonic atoms with local on-site interactions $U$ can be written as
\begin{eqnarray}
  \label{eq:model}
    H & = &\sum_i \left.\biggl[ -J\,\bigl(1+(-1)^i\delta\bigr)\,\hat a_i^\dagger \hat a_{i+1}^{\phantom\dagger} +  \hc \right. \nonumber \\
      &   & +\left.(-1)^{i}\frac{\Delta}{2} \hat n_i + \frac{U}{2} \hat n_i(\hat n_i-1)\,\right.\biggr]\,.
\end{eqnarray}
The operator $\hat a_i^\dagger$ creates a boson on site~$i$ and $\hat n_i = \hat a_i^\dagger \hat a_i$.  
Conveniently, the model can be expressed in the dimensionless quantities $U/J$, $\delta$ and $\Delta/J$. 
In the following sections, we study pump cycles which are closed cycles in the $\delta$ and $\Delta/J$ parameter space, illustrated in Fig.~\ref{fig:model}(b). 

In the absence of interactions $U=0$, the Rice-Mele Hamiltonian is a two-band model with the dispersion relation 
\begin{equation} 
  \label{eq:rm-energy}
    \epsilon_k = \pm J \sqrt{ (\Delta/2J)^2 + 4\left[1 -
    (1-\delta^2)\cos(ka)\right]}\, ,
\end{equation}
which has a degeneracy point at $\delta=\Delta=0$. This degeneracy  determines the topology of the pump cycle. If the pump path encloses the degeneracy point the corresponding pumped charge number is $\Delta Q=1$, while otherwise it is $\Delta Q=0$ [Fig.~\ref{fig:model}(b)]. 

\subsection{Phase diagram of the interacting Rice-Mele model}
\label{sec:phasediag}
\subsubsection{Hardcore-boson limit}
In the limit of hardcore interactions, $U/J=\infty$, and an average density $\bar{n}=N/L=1/2$,
with $N$ the number of bosons,
the bosonic Rice-Mele model exhibits a Mott-insulating phase, 
except for $\delta=\Delta=0$. 
This can be understood by mapping the hardcore bosons 
to a system of free spinless fermions via the Jordan-Wigner transformation \cite{Cazalilla2011}. 
At $\bar{n}=1/2$ the fermions completely fill the lowest band,
except at $\delta=\Delta=0$, 
where the system reduces to the uniform tight-binding chain with equal tunneling rates 
and the two bands merge into a single one.
In the bosonic picture, this is the point where the system becomes superfluid. 
Away from $\bar{n}=1/2$, the system is always in a superfluid phase. 
Because of this we will restrict our discussion to the average density $\bar{n}=1/2$ for the following sections.

\subsubsection{iDMRG calculations for finite $U/J$}

The 1D Bose-Hubbard model at integer filling has a gapped
Mott-insulating state at large $U/J$ and a gapless superfluid state with
algebraically decaying correlations for sufficiently low $U/J$ \cite{Bloch2008}. 
Similar physics applies to the bosonic interacting Rice-Mele
model at filling $\bar{n}=1/2$ (see, for instance, Refs.~\cite{Rousseau2006, Grusdt2013}), which also 
exhibits a superfluid-to-Mott-insulator transition.
A Mott insulator with a non-integer site occupancy is sometime referred to as a 
fractional Mott insulator \cite{Juergensen2014}.


To study the phase diagram of the interacting Rice-Mele model we use infinite-system size density matrix renormalization group (iDMRG) simulations \cite{mcculloch2008infinite} 
and compute infinite-system-size matrix-product-states (iMPS) approximating the groundstate.
This allows us to avoid edge effects and to treat the problem directly in the thermodynamic limit.
For Hamiltonians with a gapped groundstate, an iMPS approximates the groundstate very accurately. 
However, as the correlation length diverges upon approaching a critical phase,
the finite bond dimension used in the simulations imposes an effective length scale. 
In order to understand the physics independently of this effective length scale --
this is especially important in the superfluid phase --
we employ a scaling analysis in the bond dimension to elucidate the groundstate physics. 
All computations are performed with a $U(1)$ symmetry preserving code with a bond
dimension $\chi$ of up to a maximum of $\chi = 2000$ using the tensor library developed in \cite{Dorfner2016}. 
Furthermore, the local bosonic basis must be truncated in DMRG calculations to a maximum number of bosons per site. 
We found that for most of the calculations in this paper, a maximum of six bosons per site is sufficient.  

\begin{figure}
   \centering
        \includegraphics{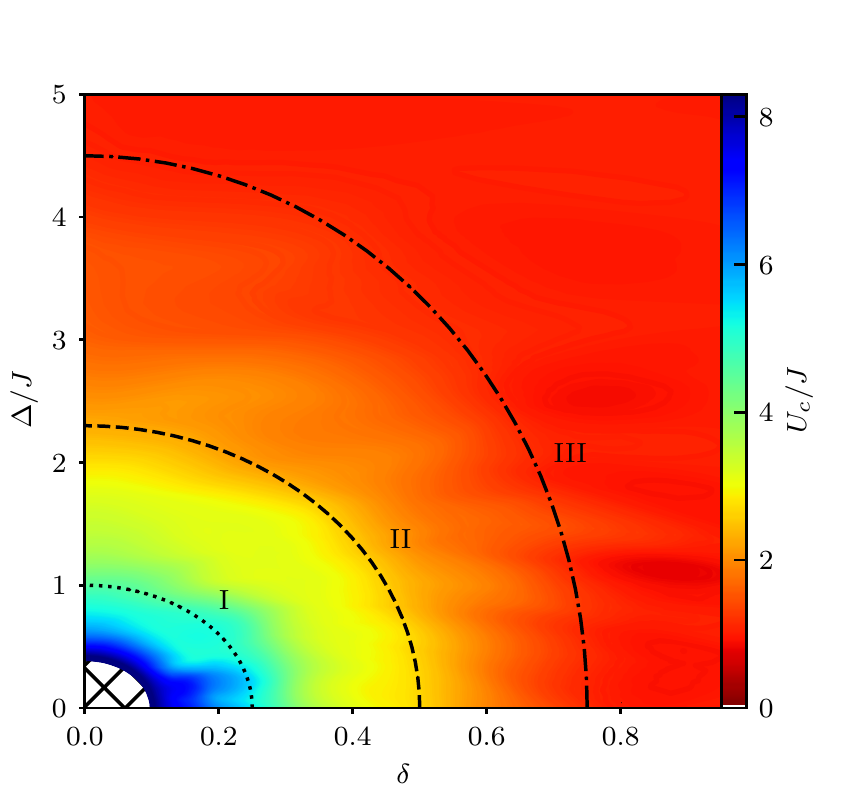}
        \caption{(Color online) {\it Phase diagram of the interacting bosonic Rice-Mele model}: Upper bound of the critical interaction strength $U_c/J$ as a 
        function of the dimensionless parameters $\delta$ (dimerization in the hopping-matrix elements) and $\Delta/J$ (staggered potential). The hatched region close to the origin has not been computed. The lines show the paths of the pump cycles studied in this paper as defined in Eq.~(\ref{eq:path-def}): dotted, path I: $R_\delta=0.25$, $R_\Delta =1.0$; dashed, path II: $R_\delta = 0.5$, $R_\Delta=2.3$; dash-dotted, path III: $R_\delta = 0.75$, $R_\Delta=4.5$. }
   \label{fig:phase-diagram}
\end{figure}

In Fig.~\ref{fig:phase-diagram}, we show a phase diagram for the bosonic
Rice-Mele model. We obtain the critical interaction strength $U_c/J$ separating
the superfluid ($U/J<U_c/J$) from the Mott insulator ($U/J > U_c/J$) as a
function of the parameters $\delta$ and $\Delta$. This diagram was computed via
a scaling analysis of the correlation length $\xi$. The superfluid-to-Mott-insulator
transition at constant density is in the universality class of
$O(d+1)$, with $d$ the dimension of the system. In our case with $d=1$, the
transition is expected to be in the same class as the Berezinskii-Kosterlitz-Thouless (BKT) transition \cite{sachdev2011quantum}, which
is the underlying assumption in our numerical analysis. Such transitions are
notoriously difficult to pin down precisely, as correlations in the system
scale as $\xi \propto \mbox{exp}(\mbox{const}/\sqrt{J/U_c - J/U})$ \cite{kuhner1998phases}, and not as a power law
like in higher dimensions. However, data from iDMRG calculations can provide an upper bound on $U_c/J$, which is sufficient for our purposes. 

The correlation lengths are extracted by taking the second largest eigenvalue
of the transfer matrix \cite{fannes1992finitely}. We find poor convergence in
the vicinity of the central degeneracy ($\Delta/J=\delta=0$), as the correlation length here diverges even in the
hardcore-boson limit. We therefore exclude  a small area
around the origin, indicated by the hatched region in
Fig.~\ref{fig:phase-diagram}.  Note that results for the phase diagram along the special lines $\Delta=0$ \cite{Grusdt2013} and $\delta=0$ \cite{Rousseau2006}
of the phase diagram were previously obtained. 

For large but finite $U/J$ the system remains gapped. One can see this by
finding the effective Hamiltonian in the single-occupancy manifold. Tracing out
 multiple occupancies results in an effective interaction:
\begin{equation}
  \label{eq:eff-int}
  H^{\mathrm{eff}}_{\rm int} = -\sum_i 4J_i^2 \left(\frac{1}{U + \Delta} + \frac{1}{U - \Delta}\right)\hat n_i\hat n_{i+1} + \mathcal{O}\left(\frac{J^3}{U^2}\right)\, .
\end{equation}
Here, we interpret the $\hat n_i = \hat b^\dagger_i \hat b_i$ as a hardcore-boson occupation
operator [with $(b_i^\dagger)^2 = 0$] and $J_i = J\left(1+(-1)^i\delta\right)$ the alternating tunneling rate. 
We switch to the equivalent   fermionic picture for an interpretation. The effective
nearest-neighbor interaction leaves the Mott-insulating phase stable and does not induce
a phase transition as long as the interaction strength is smaller than the band
gap.

\section{Charge Pumping and Many-Body Polarization}\label{sec:pump}

\subsection{Pump cycles}

\begin{figure}
	\centerline{\includegraphics{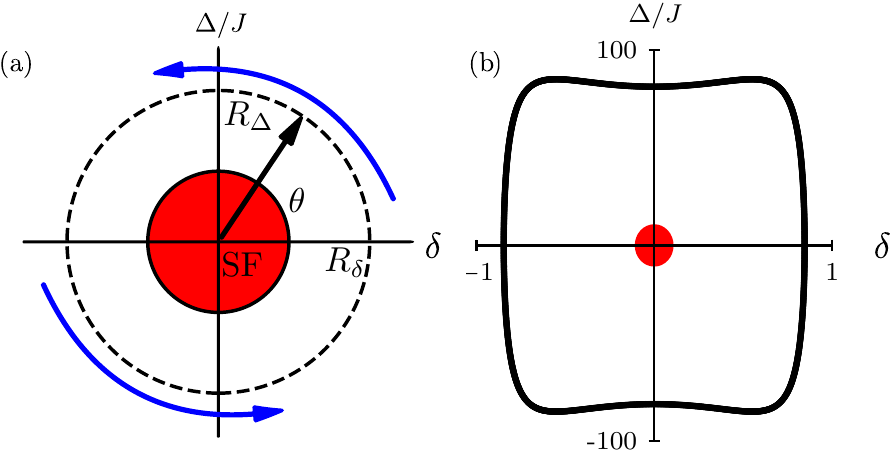}}
  \caption{(Color online) {\it Pump cycles studied in this work}: (a) Parameter space of the interacting bosonic Rice-Mele model. At the center of the parameter space, there is a superfluid region (red), which shrinks as the interaction strength increases and becomes a point as $U/J\rightarrow\infty$. We consider a charge pump that traces a path through this parameter space and encircles the superfluid phase. The cycle is parameterized by the angle $\theta$, $R_\delta$, and $R_\Delta$. (b) Pump path used in the bosonic experiment  \cite{Lohse2016} using the superlattice parameters $V_s = 10E_{r,s}$, $V_l = 20E_{r,l}$ [Eq.~(\ref{eq:superlattice})] where $E_{r,(l,s)}$ is the recoil energy of the respective lattice. The change in polarization associated with this path is shown in Fig.~\ref{fig:polarization}(b). }\label{fig:pumppath}
\end{figure}

Topological charge pumping allows for a robust quantized transport of charge
through adiabatic cyclic modulation of the system's parameters.
In the studied Rice-Mele model, pumping can be induced by an adiabatic 
modulation of the dimerization~$\delta$ and the on-site energy~$\Delta/J$.
However, it is essential that the origin of this $(\delta-\Delta/J)$ parameter space 
is enclosed by the pump path, 
as the degeneracy point at the origin is the source of the topological properties.
In this paper we focus on paths of the form
\begin{equation}
  \label{eq:path-def}
  \begin{array}{ccl}
  \Delta(\theta)/J & = & R_\Delta \sin \theta \\ \delta(\theta) & = & R_\delta \cos \theta
\end{array}
, \quad \theta \in \left[ \right.0,2\pi\left.\right),
\end{equation}
where we assume $J$ and $U$ to be constant throughout the pump cycle.
A schematic of the path is shown in Fig.~\ref{fig:pumppath}(a) while the actual paths used here 
are indicated in Fig.~\ref{fig:phase-diagram}.
We further define $\theta = 2\pi t / T$, 
assume adiabaticity and work in the instantaneous eigenbasis of $H(\theta)$. 
Hence, the charge transport becomes independent of the time scale $T$. 
 
To make a connection to experiments, we compare our results to the pump path used in the bosonic experiment~\cite{Lohse2016}, where the charge pump was implemented with a bichromatic lattice as defined in Eq.~(\ref{eq:superlattice}). 
The pump cycle in the experiment, enclosing the degeneracy point $\delta=\Delta=0$,
was realized by a variation of the relative phase~$\phi$ at constant lattice depth and is depicted in Fig.~\ref{fig:pumppath}(b).
The quantized pump was executed in the strongly-interacting regime $U\gg U_c$ and an average density $\bar{n}\simeq1/2$.

\subsection{Charge pump in the limit of hardcore bosons}\label{sec:path_hcb}
In the limit of hardcore bosons, one can compute the polarization analytically from
Eq.~\eqref{eq:pol-fermi}, using a mapping to free fermions via the
Jordan-Wigner transformation: 
\begin{equation}
P(\delta,\beta) = a\delta \beta \,\Pi \left(1-\delta^2 \big\rvert m\right),
\label{eq:pol-exact}
\end{equation}
where $\beta = \Delta/\sqrt{16J^2 + \Delta^2}$, $m = (1-\delta^2)(1-\beta^2)$, and
$\Pi$ is the complete elliptical integral of the third kind. This expression is
valid for $\delta,\beta >0$, but can be related via symmetry to the rest of the
parameter space. 

\begin{figure}
   \centering
        \includegraphics{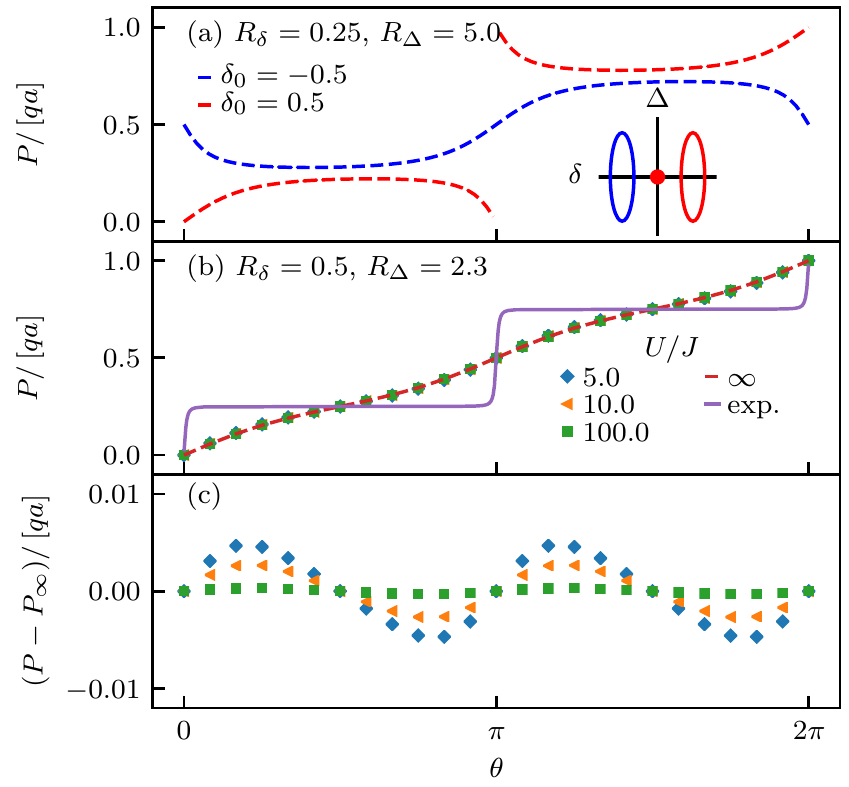}
        \caption{
        (Color online) {\it Winding of the polarization}: Polarization as determined by iDMRG
        computations [Eq.~\eqref{eq:pol-mps}] around the pump cycle with $R_\delta=0.5$, $R_\Delta=2.3$
    and different values of the interaction strengths. The charge pumped over a
    cycle is the integrated slope of the polarization [Eq.~\eqref{eq:int-current}], which simply counts the winding of the polarization. 
    (a) Polarization of hardcore bosons for paths which do not encircle the central degeneracy [see the inset in (a)],
    defined in Eq.~(\ref{eq:path-def-trivial}). Notice that the
    polarization is cyclic, but has winding number zero, implying no pumped
    charge. 
    (b)  Polarizartion $P(\theta)$ for path II ($R_\delta=0.5$, $R_\Delta=2.3$) for various interaction strengths $U/J=5,10,100,
    \infty$, and for the path used in the experiment (using $U/J=\infty$) \cite{Lohse2016}.  (c) Polarization relative  to the one in the hardcore limit, derived from the data in (b).
$P_{\infty}$ is computed from the free-fermion solution [Eq.~\eqref{eq:pol-exact}].
}\label{fig:polarization}
\end{figure}

The quantization of the pumped charge depends only on the topology of the pump path and the pumped charge is non-zero as long as the pump path encloses the degeneracy point. In the case where the path does not enclose the singularity, the polarization will have a winding number of zero, and hence no net transported charge ($\Delta Q = 0$). 
For example, consider a path defined by: 
\begin{equation}
  \label{eq:path-def-trivial}
  \Delta(\theta)/J = R_\Delta \sin \theta,  \quad   \delta(\theta) = \delta_0 + R_\delta \cos \theta \, ,
\end{equation}
where $\delta_0$ is a constant shift to $\delta$. If $R_\delta < |\delta_0|$,
then the path will not enclose the origin and the pump  will transport no
charge since, by Eq.~(\ref{eq:pol-exact}), the polarization only oscillates
around $P = 0 $ and does not wind at all, which is illustrated in
Fig.~\ref{fig:polarization}(a).  This becomes clear from  Eq.~(\ref{eq:int-current}) as the
integral of the slope of the polarization is exactly zero for the path given in Eq.~\eqref{eq:path-def-trivial}.

\subsection{Polarization from Schmidt values}
The MPS approximation to the groundstate $\ket{\Psi(\theta)}$ can be factorized via a Schmidt decomposition (used in most matrix-product states algorithms \cite{Schollwoeck2011}) into the form:
\begin{equation}
  \ket{\Psi(\theta)} = \sum\limits_{\mu=0}^{\chi-1}  \Lambda_\mu \ket{\Psi^L_\mu(\theta)} \ket{\Psi^R_\mu(\theta)}\,.
\end{equation}
The $\ket{\Psi^{L(R)}_\mu(\theta)}$ are states where $L$ and $R$ are the left and
right semi-infinite partitions of the Rice-Mele chain,  and the $\Lambda_\mu$
are the Schmidt values, which give the weight of each state.

For an infinite system, one can compute the polarization from the Schmidt
values $(\Lambda_\mu)$. The Schmidt values are also the positive root of the
eigenvalues of the reduced density matrix found by tracing out one half of the
system:
\begin{equation}
  \rho_L = \mathrm{Tr}_R\left[\rho\right] = \sum\limits_{\mu=0}^{\chi-1} \Lambda_\mu^2 \ket{\Psi^L_\mu(\theta)}\bra{\Psi^L_\mu(\theta)}\,.
\end{equation}
These numbers are directly accessible in any matrix-product-state code \cite{Schollwoeck2011}.  

The $U(1)$ symmetry associated with charge conservation allows us 
to assign an integer quantum number $\Delta N_\mu$
 to label each
 state $\ket{\Psi^L_\mu}$,  which we refer to as the
particle imbalance.  These quantum numbers are associated with particle fluctuations across
the cut in the system: The eigenstate of the reduced density matrix associated
with $\Lambda_\mu$ will have $\Delta N_\mu$ extra particles above the average density. 
Following the approach of~\cite{zaletel2014flux}, the polarization is then the expectation value of the particle imbalance: 
\begin{equation}
  \label{eq:pol-mps}
P = qa\sum_\mu  \Delta N_\mu \Lambda_\mu^2\,.
\end{equation}

The particle imbalances $\Delta N_\mu$ are only defined relative to each other.
Therefore, $P$ defined in Eq.~(\ref{eq:pol-mps}) 
is only defined modulo $qa$, as for previous definitions of the polarization.

\subsection{Charge pumping for finite $U/J$}\label{sec:pump-finite-U}
The bosonic Rice-Mele charge pump requires a finite gap  for adiabatic
transport to take place.  It is therefore necessary to avoid the superfluid
regime along the path the pump cycle takes through parameter space. 

In Fig.~\ref{fig:polarization}(b), we compare the pumped charge for a number of
paths through parameter space at finite values of $U/J<\infty$ and compare to
the hardcore-boson limit.  As expected, the polarization rises monotonically
with $\theta$. Surprisingly, the deviation $P-P_{\infty}$ [shown in Fig.~\ref{fig:polarization}(c)] of the polarization for finite $U/J$ from the $U/J\rightarrow\infty$ limit remains small, even as the correlation length in the system becomes large. These results show, from a
full many-body calculation of the polarization from Eq.~\eqref{eq:pol-MB1}, that the pumped charge remains quantized away from
the limit of hardcore bosons. This manifests itself in Fig.~\ref{fig:polarization}(c) as a vanishing difference between the polarizations of the finite and the infinite $U/J$ cases at $\theta=2\pi$. 

We would like to briefly comment on the consequence of bringing the pump
through the superfluid region. Here, the particle-fluctuations across
the system have a logarithmic divergence, with respect to the length of the
system (or, in the case of iMPS with finite bond dimensions, with respect to the
bond dimension). The relationship between particle fluctuations and entanglement
entropy has been discussed extensively in \cite{Song2012}, including the
specific case of 1D systems.  This divergence leads to an
undefined polarization for an infinite system, as the particle imbalances
$\Delta N_\mu$  in Eq.~(\ref{eq:pol-mps}) inherit this divergence. 

Similarly, the existence of gapless charge excitations means that there is, in general, no
well-defined adiabatic limit, and any measured pumped charge will depend
strongly on the details of the system and the speed with which the pump is
executed. That is to say, there is no 'universal' description of the pump in
this regime.

\section{Breakdown of the Single-Particle Interpretation for  $U/J<\infty$ in the Mott-Insulating Phase}\label{sec:mdf}
In this section, we compute the quasi-momentum distribution function of the
bosons in the hardcore-boson limit as well as for finite $U/J$ in the
Mott-insulating phase.  We show that, unless very special pump cycles are
chosen, the momentum distribution of the physical particles is never flat.
Consequently, a single-particle picture, where the total transported charge
results from the contribution of curvatures from each single-particle state,
does not lead to a quantized value for the total transported charge.

Therefore, in general, the quantization of the pumped charge in a topological charge pumping experiment with bosons 
can only be established from many-body expressions using, e.g., the polarization [Eq.~\eqref{eq:pol-MB1}]. 
We argue that the breakdown of a single-particle topological invariant could be demonstrated by measuring 
the quasi-momentum distribution of each band and showing that the integral over this $n_\alpha(k, \theta)$ and the known single-particle 
Berry curvature is not quantized.

\subsection{Quasi-momentum distribution function}\label{sec:qmdf}
\begin{figure}
   \centering
        \includegraphics{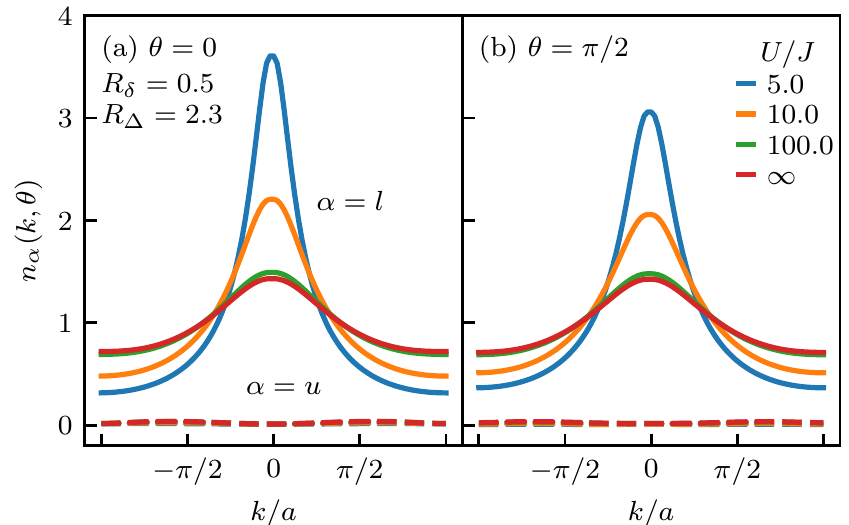}
        \caption{(Color online) {\it Quasi-momentum distribution of the interacting system}: $n_{\alpha}(k, \theta)$ for
        the pump cycle with $R_\delta=0.5$, $R_\Delta=2.3$. Solid lines are the
    projection onto the lowest band ($\alpha=l$) of the noninteracting
Hamiltonian, the dashed lines are for the projection onto the upper band
($\alpha=u$). (a) Distribution at $\theta=0$. (b) $\theta = \pi/2$. For
the path shown, the occupation density of the upper band is quite low:
$n_u(k,\theta) \lesssim  0.03$ in the $U/J=\infty$ limit for all $\theta$. However,
this occupation can be larger when the pump cycle is closer to the degeneracy.
Note  that for the experimental path of Ref.~\cite{Lohse2016} as shown in Fig.~\ref{fig:pumppath}, the momentum distribution is essentially
flat and confined completely to the lower band.} \label{fig:n_of_k}
\end{figure}

In order to compute the quasi-momentum distribution function, we need to
compute the one-particle density matrix (OPDM). It is straightforward to
extract the OPDM from iMPS solutions using transfer-matrix methods
\cite{eisert2013entanglement}.


The OPDM is given 
by $\rho^{(1)}_{ij} = \braket{\Psi|\hat{a}^\dagger_i
\hat{a}_j|\Psi}$, where $|\Psi \rangle$ is, in our case, the many-body groundstate. 
For a translationally
invariant state, the corresponding OPDM $\rho^{(1)}$ will share the symmetry 
and can be block-diagonalized into blocks of dimensions $n \times n$, where $n$
is the size of the unit-cell, and each block can be labeled by a
quasi-momentum.  

The quasi-momentum distribution usually measured in an  experiment results from the projection of the
OPDM onto the bands of a noninteracting model \cite{Bloch2008}. This corresponds to taking the diagonal elements
of $\rho^{(1)}(k)$ when written in the eigenbasis of the noninteracting model.
One can then associate a momentum distribution $n_{\alpha}(k,\theta)$ with each original band ($\alpha=1,\dots,n$ is the band index), as
shown in Fig.~\ref{fig:n_of_k}.

In the present case, we have a two-site unit-cell, and we refer to the
lower(upper) bands as $\alpha = l(u)$.  In experiments, a flat $n_\alpha(k,\theta)$ is
usually achieved by either using free fermions and choosing the filling
appropriately or by localizing particles into individual sites.  
In the case of hardcore bosons, they, in general,
already lack a perfectly flat momentum distribution for any filling of less than one
particle per site and in equilibrium. Only after the relevant Jordan-Wigner
transformation (see, e.g., \cite{Gangardt2006}) the result is a flat quasi-momentum distribution, but for spinless fermions.  However, in a
sufficiently deep lattice, the bosons will be largely localized, with an essentially
flat distribution, which was exploited in \cite{Lohse2016} (see also our discussion
in Sec.~\ref{sec:path_hcb}).

In Fig.~\ref{fig:n_of_k}, the quasi-momentum distribution is plotted for two points $\theta=0,\pi/2$
along the pump path with $R_\delta=0.5$ and $R_\Delta=2.3$ for a number of
interaction strengths, including the hardcore-boson limit ($U/J=\infty$). The
projection onto both the upper ($\alpha=u$) and the lower band ($\alpha=l$) of the
noninteracting model is shown. The momentum distribution of the physical particles
for this path is far from the flat, fermionic distribution, even in the hardcore-boson
limit. 
As the interaction strength approaches $U_c$, the
momentum distribution becomes increasingly peaked  at $k=0$ and develops power-law tails.

We find that  the projection onto the upper band is relatively small [$n_u(k,\theta) \lesssim 0.03$], even far
away from the hardcore limit, although it increases as one approaches the central
degeneracy.   

\subsection{Integrated single-particle Berry curvature}
\begin{figure}
   \centering
        \includegraphics{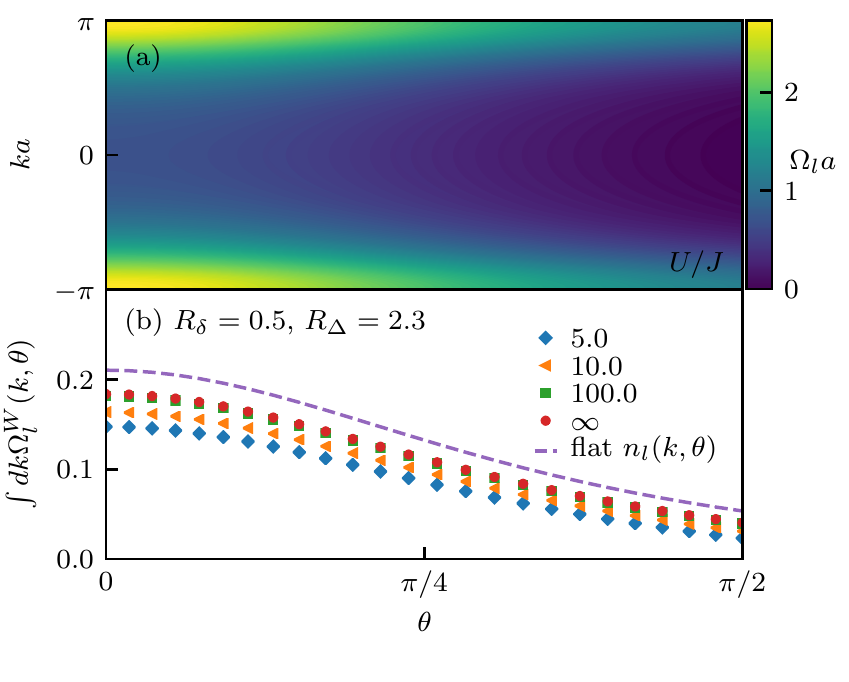}
	\caption{(Color online) {\it Single-particle Berry curvature}:
(a) Berry curvature $\Omega_l(k,\theta)$ [Eq.~(\ref{eq:berry-curvature})]
        used for computing the integrated curvature.
		(b) Partially integrated weighted curvature
		[defined in Eq.~(\ref{eq:alt-int-curve})], along the pump cycle for path II. The total integrated
		charge $\Delta Q'_l/q$ after one pump cycle is simply the area
		under these curves between $\theta = 0$ and $2\pi$. 
		 $\Delta Q'_l/q$ for all paths can be seen in Fig.~\ref{fig:delta-q} and are smaller 
		than one.  }\label{fig:integrated}
\end{figure}

The expression Eq.~\eqref{eq:chern-number} for the pumped charge $\Delta Q$ for
noninteracting fermions can be interpreted as the accumulated curvature picked
up in each filled band over the course of a pump cycle.  In the interacting
case, one might, by analogy, consider the momentum-weighted single-particle
Berry curvature, which we define as 
\begin{equation} \label{eq:alt-int-curve} \Omega^W_\alpha(k, \theta) =
\Omega_\alpha(k, \theta)n_\alpha(k, \theta), \end{equation}
with the single-particle Berry curvature and the momentum distribution for each
band~$\alpha$.  The maxima of the single-particle Berry curvature in $k$-space
occur at the band-gap minima where $ka=\pm\pi$; along the pump path they occur at
$\theta = m \pi$ for $m \in \mathbb{Z}$, but it still has a significant
magnitude throughout the pump cycle.  
Therefore, charge is pumped at any point during the cylce. 
Figure~\ref{fig:integrated}(a) shows the
single-particle  Berry curvature $\Omega_l$ for the lower band of path II ($R_\delta = 0.5$,
$R_\Delta=2.3$).

The single-particle picture which motivates Eq.~(\ref{eq:alt-int-curve})
assumes that we simply sum up the contributions from each momentum state to compute the
transported charge, which leads to a quantized $\Delta Q_\alpha$ when there is a filled
band.  However, for our model with a finite $U/J$, this will no longer
necessarily give a quantized amount for the charge transport due to the inhomogenous occupation in $k$ space.

In analogy to Eq.~(\ref{eq:chern-number}), we can define the integral of the  weighted single-particle Berry curvature
$\Omega^W_\alpha$ defined in Eq.~\eqref{eq:alt-int-curve}:
\begin{equation}
\label{eq:qprime}
\Delta Q'_\alpha = q\int^{\pi/a}_{-\pi/a} dk \, \int_0^{2\pi} d\theta\, \Omega^W_\alpha(k,\theta).
\end{equation} 
$\Delta Q_\alpha'$ will, in general, deviate from the exactly quantized $\Delta
Q_\alpha$ since, in general, $n_\alpha (k, \theta)$ is not flat. 

\begin{figure} 
  \centering
  \includegraphics{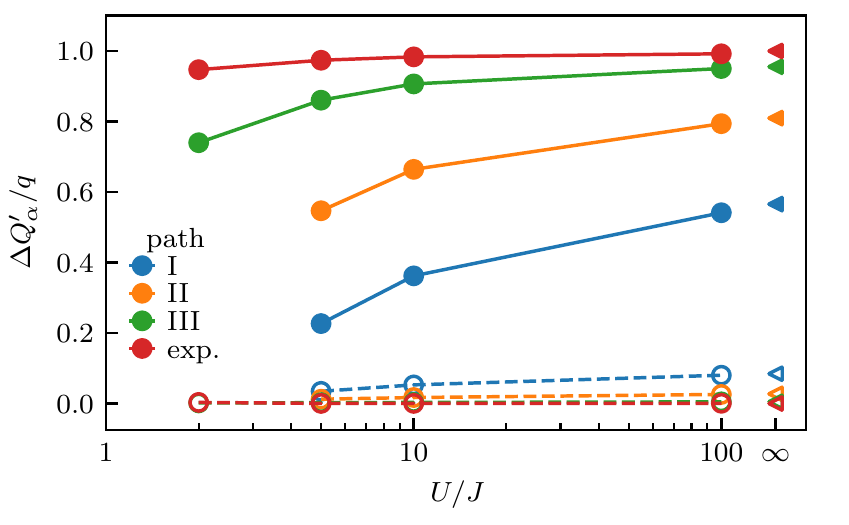}
  \caption{(Color
    online) {\it Breakdown of quantized single-particle invariant}: Integral  $\Delta Q'_\alpha/q$ over the weighted single-particle Berry curvature [defined in
    Eq.~(\ref{eq:qprime})] after one pump cycle. 
We also include the path used in the bosonic charge-pump experiment \cite{Lohse2016} for comparison, for which the numbers are very close to 1.
In general, though,    $\Delta Q'_\alpha$
    is  not quantized. Full(empty)
symbols indicate $\Delta Q'_\alpha$ for the lower(upper) band. Triangles indicate the
$U/J\rightarrow\infty$ limit. 
} \label{fig:delta-q}
\end{figure}

The partially integrated weighted curvature is shown in Fig.~\ref{fig:integrated}(b) for different points during the pump cycle. The contributions are generally smaller than the corresponding values obtained for a filled band of noninteracting fermions  [dashed line in Fig.~\ref{fig:integrated}(b)]. 
Results for  $\Delta Q'_\alpha$ are plotted in Fig.~\ref{fig:delta-q} as a function of $U/J$ for the paths shown in
Fig.~\ref{fig:phase-diagram} and for the experimental path.
For the experimental path, $\Delta
Q'_l\approx q$ even for small interaction strength. 
The reason is  that the bosons remain essentially localized to individual sites during the entire pump cycle.
For the elliptical paths that take the system closer to the superfluid region and even for hardcore bosons, $\Delta Q'_l/q$ can deviate significantly from one. For instance, for path II at $U/J = \infty$, we have $\Delta Q'_l/q = 0.81$.
This is consistent with the fact discussed above (see Sec.~\ref{sec:qmdf})  that the momentum distribution of hardcore bosons is, in general,
not flat for any filling of less than one boson per site.

\section{Aspects of an Experimental Realization}\label{sec:exp}
Finally, we discuss deviations from ideal conditions  relevant to an experimental realization of interacting charge pumps.
These aspects are of similar nature as those already found in previous experiments~\cite{Lohse2016,Nakajima2016}.
In particular, the transport may be affected by  imperfect initial state preparation, trap and finite-size effects.
Furthermore, a high measurement accuracy 
is  necessary to resolve the deviations in the transport.
However, the main challenge is to perform the interacting charge pump adiabatically
with respect to the many-body gap,
while keeping technical heating processes sufficiently low.
In the hardcore-boson limit, the relevant gap is given by the minimum single-particle band gap
[see  Eq.~\eqref{eq:rm-energy}], which corresponds to $\Delta E/J = 1$ for pump path II.
When moving away from the hardcore interacting case, the many-body gap reduces rapidly. For example, 
the minimal many-body gaps for pump path~II at two exemplary interaction strengths are 
$\Delta E/J=0.075$ for $U/J=5$ and $\Delta E/J=0.21$ for $U/J=10$.
From Fig.~\ref{fig:delta-q}, we know 
that for this pump path,  significant deviations from the single-particle description exist even in the hardcore-boson limit,
which we expect to be observable in future experiments by combining the center-of-mass measurements \cite{Lohse2016,Nakajima2016} with observations of the momentum distribution $n_\alpha(k,\theta)$ using band-mapping techniques~\cite{Bloch2008}.

\section{Entanglement Spectrum}\label{sec:ent}
The entanglement spectrum,
defined with respect to a spatial partition of the system into two (semi-infinite)
halves, is given by the eigenvalues of the entanglement Hamiltonian $H_E$ \cite{li2008entanglement}:
\begin{equation}
	\rho_L = e^{-H_E} \,.
\end{equation}
$H_E$ is unitless and is constrained to be strictly positive by the
normalization requirement of the reduced density matrix. The eigenvalues of
$H_E$ are referred to as entanglement eigenvalues (EEVs).

\begin{figure}
   \centering
        \includegraphics{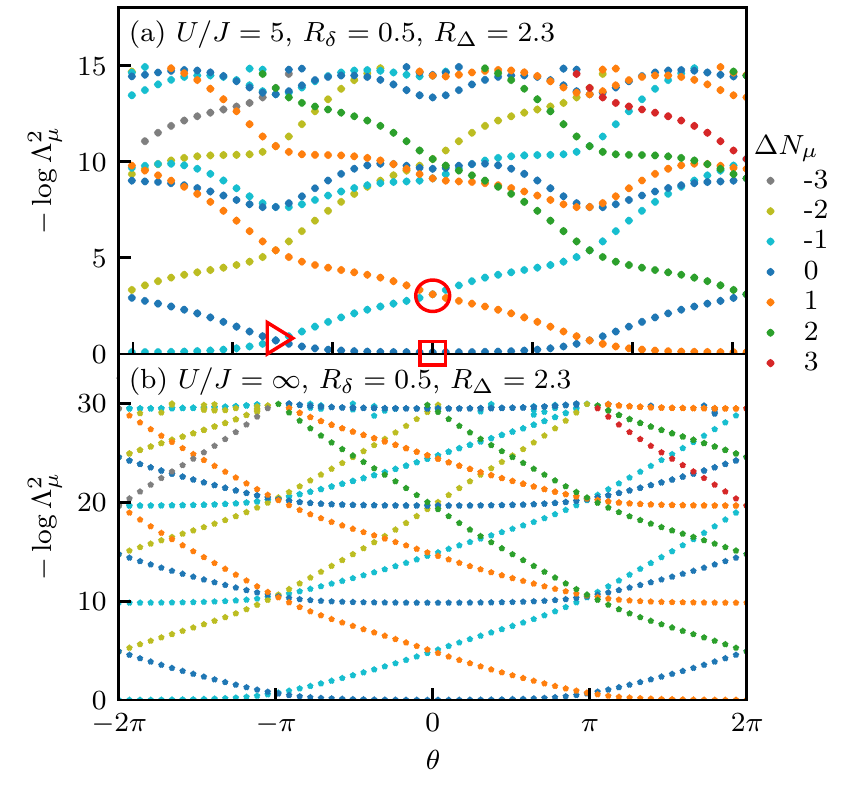}
        \caption{(Color online) {\it Entanglement spectrum of the Rice-Mele model}: Results for the pump cycle with
          $R_\delta=0.5$, $R_\Delta=2.3$ for  (a) $U/J = 5$ (from iDMRG) and  (b)  the
          $U/J=\infty$ limit, the latter calculated from the free-fermion solution. Note
          the different scale for the two spectra.
     The labels $\Delta N_\mu$ correspond to the particle imbalance for each Schmidt value. Notice that
   entanglement eigenvalues in the spectrum  wind either up or down or have no winding  if $\Delta N_\mu =  \Delta N_0$. After one cycle, the spectrum has the same values, but with all labels increased by 1, indicating the pumping of a  single charge.}
   \label{fig:schmitt}
\end{figure}

The flow of the  entanglement spectrum is smooth and continuous for a gapped state which
evolves under adiabatic perturbations, as is the situation in the charge
pump under consideration here. 
The topological nature of the charge pump is revealed directly in the spectral
flow of the entanglement spectrum.  Figure~\ref{fig:schmitt} shows the
entanglement spectrum over the course of one pump cycle ($R_\delta=0.5$, $R_\Delta=2.3$).  Particle-number conservation allows
each EEV  to be labeled with a  particle imbalance $\Delta N_\mu$ across the
chosen cut. 
We plot the EEVs for    $-2\pi<\theta < 2\pi $ and first focus the discussion on the interval $\theta \in \left[-\pi, \pi \right)$, which results in a unique lowest EEV ($\mu=0$)  for $\theta \in \left(-\pi, \pi \right)$, to which we assign the imbalance $\Delta N_0=0$. 

	Figure~\ref{fig:schmitt}(a) shows the spectrum for $U/J= 5$, computed using iDMRG. Using matrix-product states to represent the ground state throughout the pump cycle
makes studying entanglement spectra particularly easy as the eigenvalues $\Lambda_\mu$ of the reduced density matrix are computed as part  of the iDMRG calculation along
with the $\Delta N_\mu$. Figure~\ref{fig:schmitt}(b) shows the corresponding data for $U/J=\infty$, computed from the free-fermion solution (see, e.g., \cite{Peschel2009}). 

First, note the two symmetric points at $\theta = 0$ and at $\theta = -\pi$
($\pi$), corresponding to the trivial and non-trivial phases of the
SSH model, respectively. This arises from the lattice inversion symmetry and imposes a
structure on the spectrum. In general, this protected symmetry also requires a fermion parity symmetry~\cite{sticlet2014fractionally}, but in our model this is automatically satisfied by the $U(1)$ symmetry.  The inversion symmetry implies that the spectrum
should be invariant under $\Delta  N_\mu\leftrightarrow -\Delta N_\mu$. This
requires that EEVs must come in pairs with $\Delta N_\mu$, $\Delta N_\nu$ ($\mu\not=\nu$) such that  $\Delta N_\mu +
\Delta N_\nu = N_0(\theta)$ or  as singlets with $\Delta N_\mu = N_0(\theta)$, where $N_0(\theta)$
is a (integer) constant that can differ for  $\theta\in \lbrace -\pi ,0 \rbrace$. The existence or absence of the singlet (or, equivalently, if $N_0$ is even or odd) distinguishes
between the two phases of the SSH model. 
This relationship between symmetry and entanglement spectrum is well established \cite{Turner2011,Sirker2014}. 

Furthermore,  the spectrum is periodic but  the different EEVs can have
non-trivial winding over a pump cycle.  EEVs from a pair  at $\theta=-\pi$ with
$\Delta N_{\mu}<0$ ($\Delta N_{\mu}>0$)  wind up (down) and EEVs with $\Delta N_\mu =
0$ have trivial winding. After one cycle, the spectrum has the same set of EEVs,
but with $\Delta N_{\mu} \rightarrow \Delta N_{\mu}+1$, increased by one [compare
the imbalances for $|\theta| >\pi$ to those for $\theta \in (-\pi ,\pi)$].

The increase of labels can be understood by considering the spectral flow
between the topologically non-trivial SSH state (at $\theta = -\pi$) and the trivial one ($\theta = 0$). At $\theta = -\pi$, there
is no singlet but only pairs. To reach the trivial phase at $\theta=0$, 
pairs containing a state with $\Delta N_\mu=0$ must split up, with one state 
becoming a singlet at $\theta=0$, and the other forming a pair with another EEV
from a different pair. This is marked in Fig.~\ref{fig:schmitt}(a) as squares
and circles, respectively.  For example, in our choice of $\Delta N_\mu$,  at
$\theta=0$ we have $N_0=0$, and the singlets (red square) have $\Delta  N_\mu=0$.
The singlet is connected to pairs at $\theta =-\pi$ of $\Delta  N_\mu= 0,-1$ (red
triangle) and $N_0=-1$. The EEV with $\Delta  N_\mu= 0$ becomes the singlet and
the EEV with $\Delta  N_\mu= -1$ forms a pair with an EEV with $\Delta  N_\mu= +1$
(red circle). 

At $\theta = \pi$, the EEVs must all form pairs again. The non-trivial winding
of the EEVs implies that $N_0(\pi) \not= N_0(-\pi)$. In our case, $N_0(\pi)=1$,
and the lowest pair now has $\Delta  N_\mu= 0,1$ as discussed above.

The connection to charge pumping can be made explicit by considering the
expression for the many-body polarization in Eq.~(\ref{eq:pol-mps}). The shift in
$\Delta N_\mu$ by some integer over one cycle implies that the polarization
changes smoothly and also increases by exactly the shift:
\begin{eqnarray}
	P(\theta + 2\pi) &=& qa\sum\limits_\mu \Lambda_\mu^2(\theta + 2\pi) \Delta  N_\mu \nonumber\\
	 & =& qa\sum\limits_i \Lambda_\mu^2(\theta) (\Delta N_\mu + 1) \nonumber \\
	&=& P(\theta) + qa\, .
\end{eqnarray}
Finally, we compare the entanglement spectra at $U/J=5$ to the one at $U/J=\infty$ [see Figs.~\ref{fig:schmitt}(a) and (b)].
As $U/J$ decreases, the EEVs shift to lower values, thus compressing the spectrum and giving
more weight to higher particle fluctuations. Moreover, the entanglement
spectrum of hardcore bosons exhibits higher degeneracies, inherited from the
free-fermion case. These degeneracies are broken by the interaction term in
Eq.~(\ref{eq:eff-int}), since in general, the states with higher particle
fluctuations have lower particle density, resulting in a lower interaction energy for these states. The topological structure described above, however, is preserved.

\section{Summary}\label{sec:sum}
We  showed that topology protects  
the quantized charge pumping in the interacting bosonic Rice-Mele model in the Mott-insulating regime of one boson per unit cell  away from the
regime of hardcore bosons. The computation of the quantized pumped charge requires a full many-body calculation via the polarization. We further demonstrated that for interacting bosons with $U_c/J<U/J \leq\infty$, the single-particle properties of the physical particles never 
capture the quantized pumping. In the hardcore-boson limit, due to  the mapping to free fermions via the Jordan-Wigner transform, a single-particle
interpretation is still possible in terms of these fermions. 
We propose to carry out an experiment in the Mott-insulating regime but with $U/J<\infty$. 
A measurement of the center--of-mass of the cloud 
would establish the quantized charge per pump cycle. Measuring the quasi-momentum distribution throughout the pump cycle would allow
one to demonstrate the breakdown of a single-particle interpretation when combining this information with the 
known single-particle Berry curvature of the bands.

We also show how the  quantized transport is reflected in the structure of the
entanglement spectrum and the symmetries of the lattice. The existence of
different symmetry-protected topological phases inherited from the SSH model along the pump cycle constrains 
the degeneracy structure of the entanglement spectrum, enforcing the
quantization of the charge transport in the Mott-insulating regime.  
\\

We acknowledge very helpful
discussions with I. Bloch, V. Gurarie and F. Pollmann.
This work was supported by the Deutsche Forschungsgemeinschaft (DFG, German
Research Foundation)  under  project number 277974659 via Research Unit FOR 2414. 
C.S., M.L. and M.A.
additionally acknowledge support from the DFG - project number 282603579 (DIP),
the European Commission (UQUAM Grant No. 5319278), and the Nanosystems
Initiative Munich (NIM) Grant No. EXC4.  
The idea for this
project was conceived at  KITP, University of Santa Barbara,  during the
program SYNQUANT16 in discussions with I. Bloch. The hospitality at KITP is
gratefully acknowledged.  This research was supported in part by the National
Science Foundation under Grant No. NSF PHY-1748958.

\bibliographystyle{new-apsrev}
\bibliography{References}
\end{document}